\documentclass[aps,prx,twocolumn,final,letterpaper,floatfix,superscriptaddress]{revtex4-2}
\usepackage{appendix}
\usepackage{graphicx}   
\usepackage{import}
\usepackage{comment}
\usepackage{epstopdf}
\usepackage{bm}
\usepackage{amssymb}
\usepackage{quotes}
\usepackage{braket}
\usepackage{transparent}
\usepackage{dcolumn}
\usepackage{multirow}
\usepackage{mdframed}
\usepackage{booktabs}
\usepackage{color}
\usepackage{bm}
\usepackage{dsfont}
\usepackage{slashed}
\usepackage{enumitem}
\usepackage{amsmath} 
\usepackage{changepage} 
\usepackage{hyperref}

\hypersetup{
    colorlinks=true,
    citecolor=blue, 
    linkcolor=blue, 
    urlcolor=blue   
}

\begin{document}
\rmfamily

\title{Diffractive cascades for polychromatic hard X-ray focusing}

\author{William Michaels}
\email{willmich@mit.edu}
\affiliation{Research Laboratory of Electronics, MIT, Cambridge, MA 02139, USA}
\affiliation{Department of Electrical Engineering and Computer Science, MIT, Cambridge, MA 02139, USA}

\author{Simo Pajovic}
\affiliation{Department of Mechanical Engineering, MIT, Cambridge, MA 02139, USA}

\author{Joshua Chen}
\affiliation{Research Laboratory of Electronics, MIT, Cambridge, MA 02139, USA}
\affiliation{Department of Electrical Engineering and Computer Science, MIT, Cambridge, MA 02139, USA}

\author{Charles Roques-Carmes}
\affiliation{Research Laboratory of Electronics, MIT, Cambridge, MA 02139, USA}
\affiliation{Edward L. Ginzton Laboratory, Stanford, CA 94305, USA}

\author{Marin Solja\v{c}i\'{c}}
\email{soljacic@mit.edu}
\affiliation{Research Laboratory of Electronics, MIT, Cambridge, MA 02139, USA}
\affiliation{Department of Physics, MIT, Cambridge, MA 02139, USA}
\begin{abstract}

Diffractive focusing of hard X-rays has traditionally required structures with large aspect ratios due to the limited interaction of most materials with X-rays. This has increased the complexity of fabricating diffractive X-ray lenses, restricting their widespread deployment. Here, we utilize topology optimization to design diffractive cascades to focus X-rays. When restricting the structures to a maximum aspect ratio of 8, a diffractive cascade can achieve a focusing efficiency of 40\%, far exceeding the 3\% efficiency of a zone plate with the same aspect ratio. Diffractive cascades also allow the focusing of beams with energies beyond 20 keV and bandwidths exceeding 1\%, loosening the restrictions on other system components. We characterize the robustness of these cascades to alignment, fabrication, and heating perturbations, demonstrating the ability of our designs to operate under real-world conditions. Finally, we exploit the flexibility of our framework to include multiple depths in the objective function. This enables a depth of focus exceeding that of a zone plate or a cascade designed using single-plane optimization. This work demonstrates the utility of topology optimization in the X-ray regime and the possibility of advancing X-ray manipulation across a range of tasks. 
\end{abstract}
\maketitle


\section{\label{sec:level1}Introduction}
Wavefront control of X-rays is difficult due to their weak interactions with most materials. This same property gives X-rays much of their utility, as their penetrative ability means they can be used to image structures within objects that are invisible to lower energy photons. Moreover, the short wavelengths of X-rays mean that, in principle, they can be used to image nanoscale structures beyond the diffraction limit of visible photons. Despite this, the highest resolution X-ray imaging techniques are limited by the size of the X-ray focal spot on the sample. Therefore, improved X-ray focusing would enable enhanced imaging capabilities across applications such as materials science, medicine, biology, and archaeology \cite{sakdinawat2010nanoscale}. 

\begin{figure*}
    \centering
    \includegraphics[scale=0.54]{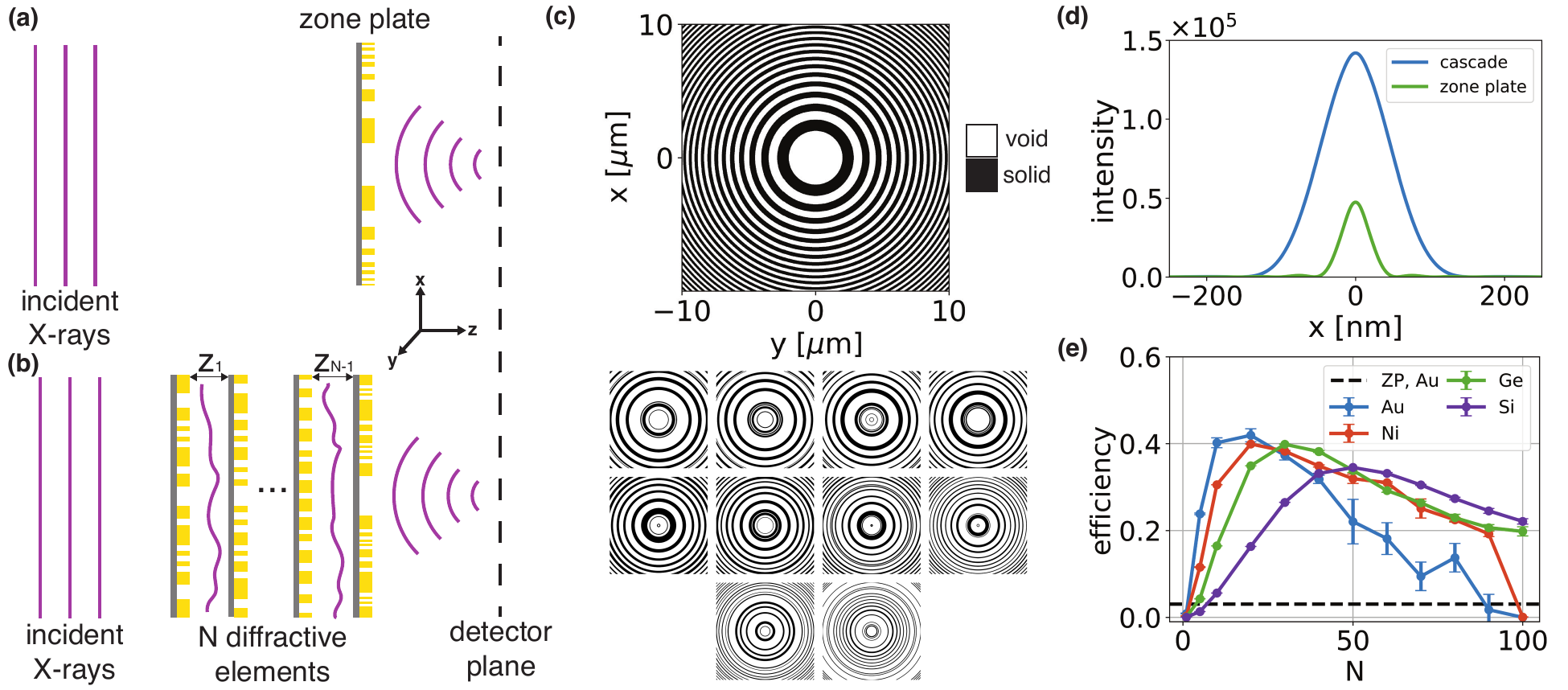}
    \caption{\textbf{Diffractive cascades for X-ray focusing.} \textbf{(a)} Schematic diagram of X-ray focusing with a zone plate. \textbf{(b)} Schematic diagram of X-ray focusing with a diffractive cascade. Each pixel of each element in the cascade is an optimization parameter, and each element has a Si$_3$N$_4$ membrane attached. The focal length is defined as the distance from the last element to the focal plane. \textbf{(c)} Comparison of a single zone plate and cascade elements. The first cascade element is in the top left, with subsequent elements being read off from left to right and top to bottom. Note that only a small region in the center of the zone plate and cascade elements is shown for legibility. The elements considered here have a diameter of 82 $\mu$m. \textbf{(d)} Focal spot comparison between zone plate and cascade for design shown in (c). The cascade achieves a similar spot size and higher intensity across the focal spot. The input wavefront was normalized to have a uniform intensity of 1. \textbf{(e)} Scaling of focusing efficiency with the number of elements in the cascade, $N$. For materials with lower loss we see a higher optimal number of elements compared to materials with higher loss. For each combination of material and number of elements, multiple optimizations with different random parameter initializations are performed. The results of these optimizations are used to compute the mean and error bars for each combination in the plot. ZP, Au refers to the efficiency of a gold zone plate restricted to the same aspect ratio as the elements in each cascade.} 
    \label{fig:1}
\end{figure*}

Traditionally, X-ray focusing has been accomplished using optical components such as compound refractive lenses \cite{snigirev1996compound, lengeler1999microscope}, grazing incidence mirrors \cite{yumoto2020nanofocusing, takeo2020soft, yamada2024extreme}, polycapillaries~\cite{macdonald2003applications}, zone plates \cite{lai1992hard, yun1999nanometer, david2011nanofocusing}, multilayer Laue lenses \cite{bajt2018x, huang201311}, or combinations of these elements \cite{snigirev2007submicrometer, doring2013sub}. Among the most popular and elegant X-ray lenses are zone plates (Figure \ref{fig:1}(a)), which are circularly symmetric elements designed according to an analytical formula \cite{soret1875phenomenes}. Zone plates have achieved impressive performance across multiple metrics. Stacks of interlaced zone plates have achieved sub-10 nm spot sizes for X-ray energies of 9.1 keV \cite{mohacsi2017interlaced}. However, due to their sub-$\mu$m thicknesses, the focusing efficiency of these devices (defined as the proportion of power focused into the first diffraction order) is less than 1\%.

Fundamentally, the focusing efficiency of zone plates is strongly dependent on increasing their thickness and decreasing their minimum feature size. Theoretically, the efficiency of a zone plate is maximized when it imparts a $\pi$ phase on the incoming light. Unfortunately, for most materials, this condition implies thicknesses exceeding 2 $\mu$m for hard X-rays. Combined with desired minimum feature sizes as small as $\sim$50 nm, this necessitates aspect ratios over 40, which are challenging to fabricate. One way this problem has been tackled is by creating bespoke fabrication techniques to increase the achievable aspect ratio. Specifically, novel resist chemistries and stabilizing polymer molds have been developed towards this end \cite{spector1997process, feng2007nanofabrication}. Such methods have pushed zone plate efficiencies beyond 5\% in the 6-10 keV range \cite{li2020tunable, mohacsi2015high}. Another way to scale the aspect ratio of zone plates is through cylindrical fabrication. Instead of using a planar substrate, cylindrical multilayer zone plates are fabricated by sputtering material on a spinning rod and then cutting the cylinder to the desired length \cite{rudolph1982status}. Such devices have been demonstrated focusing 20 keV X-rays with 20\% efficiency \cite{takano2017hard}. This cylindrical fabrication removes etching and lithography constraints but introduces complexity and can create zone uniformity issues.  

Thus, substantial performance gains have been realized by keeping unchanged the original zone plate formula but improving its fabrication. However, such high aspect ratio structures can create mechanical instabilities and suffer from irreproducibility due to the highly specific nature of their fabrication methods. Furthermore, these structures introduce parasitic volume diffraction effects that lower performance \cite{schneider2008volume}. Hence, other work has been done to modify the zone plate structure to improve performance \cite{maser1992coupled, mohacsi2014high}. However, these investigations have been limited to tuning only a few parameters or otherwise significantly restricting the form of the diffractive elements. Investigations of small cascades of X-ray zone plates have been published \cite{cagniot2010cascades, gleber2014fresnel, rehbein2015near, wojcik2015stacking, mohacsi2017interlaced}, but they are unable to scale to many elements due to the resulting increase in design complexity. More generally, ``diffractive cascades'' containing elements with arbitrary geometries have been designed for imaging or computing tasks, but they have been limited to visible wavelengths where material and aspect ratio restrictions are much more forgiving \cite{gulses2013cascaded, ferdman2022diffractive, motz2025design}. 

Due to the complex tradeoffs within X-ray focusing and the large room for improvement over zone plates, photonic inverse design is a well-suited approach to this problem. Here, we use topology optimization, whereby the optical element is parameterized by the material density at each point on the element. Topology optimization has been applied in photonics to a diverse set of tasks and has enabled performance gains over previous photonic designs for various applications \cite{piggott2015inverse, shen2015integrated, raza2024fabrication, efseaff2023strategies, udupa2019voxelized}. Modern topology optimization techniques are general and can be applied successfully to problems with thousands or millions of parameters \cite{jensen2011topology}. Despite the ubiquity of these methods, there have been few applications of topology optimization to the design of X-ray optics \cite{lee2023direct}. This is partially due to the widespread assumption that photonic inverse design requires wavelength-scale features. Such features are intractable to simulate and fabricate in the X-ray regime. However, diffractive X-ray optical elements operate with features orders of magnitude larger than the incident wavelength and can be parameterized in the same way as visible wavelength structures. Furthermore, the low numerical aperture of diffractive X-ray lenses and the cylindrical symmetry of X-ray focusing allows for the use of a 1D scalar diffraction forward model, greatly speeding up optimization when compared to techniques like vectorial finite-difference time-domain solvers. 

In this work we apply topology optimization to the design of cylindrically symmetric binary X-ray lenses. These lenses cascade multiple diffractive elements together (Figure \ref{fig:1}(b)) and result in high-efficiency focusing of hard X-rays at low aspect ratios. These cascades can operate at higher energies and larger bandwidths than traditional zone plates. They are also robust to alignment and fabrication errors, as well as thermal expansion due to heating from excessive X-ray flux. The objective function can be easily modified to consider other figures of merit such as the depth of focus for capabilities beyond those of analytical elements. These results open the possibility of fabricating high performance diffractive X-ray lenses without the use of specialized fabrication techniques. It also suggests the utility of inverse design principles in the X-ray regime where they have rarely been used. 

\section{Topology optimization of diffractive cascades}
We propose improving upon traditional zone plates by creating a cascade of diffractive elements between the source and the focal plane. Each of the elements is circularly symmetric, aligned perpendicular to the optical axis, and parameterized by the density profile of material along its radius. We benchmark our design against a zone plate whose zone locations are given by the equation~\cite{booklet2001x}
\begin{equation}\label{eq:1}
    r_n = \sqrt{n\lambda f + \frac{n^2\lambda^2}{4}},
\end{equation}
where $n$ is the zone number, $f$ is the focal length, and $\lambda$ is central wavelength of illumination. Schematics showing the difference between a zone plate and an example diffractive cascade is shown in Figure \ref{fig:1}(a)--(c). A key aspect of Eq. \ref{eq:1} is that it is derived for a single wavelength, indicating that zone plates are typically designed to be narrowband. Our topology optimization approach is agnostic to the input spectrum, allowing for optimization over a broad range of wavelengths.

The forward model assumes polychromatic uniform illumination incident on the focusing structure. The incident waves are assumed to be fully spatially coherent with a uniform intensity across their wavefront. The illumination is centered at a central energy with a finite bandwidth to reflect conditions found in a typical synchrotron or free-electron laser beamline. The resulting focal spot of the structure is calculated using scalar diffraction between elements and the thin element approximation at each element. The objective function is then computed as the power in a small region in the center of the focal plane. To speed up the optimization, the forward model propagation is computed along a 1D radial slice and the structures are revolved around the optical axis at the end of the optimization to create the full 2D, cylindrically-symmetric elements. We then use a 3D scalar optics simulation to validate the design and compute the final intensity distribution. We perform all computations on a GPU to further reduce the runtime of the optimizations. Since the numerical apertures of all the elements considered here are $\sim 10^{-3}$, it is a reasonable approximation to not consider the full vector version of Maxwell's equations \cite{goodman1969introduction}. 

The optimization of the cascade is performed using the NLopt software package \cite{NLopt}. Specifically, we use the NLopt implementation of the method of moving asymptotes \cite{CCSA}, a gradient-based optimization algorithm that performs well for topology optimization problems with many parameters. Gradients are computed using automatic differentiation in PyTorch \cite{paszke2019pytorch}. To ensure a binary design at the end of the optimization procedure, we use a thresholding procedure to gradually binarize the parameters. Following the standard topology optimization framework, at the start of the optimization the density of material at each point is allowed to vary continuously between 0 and 1 \cite{jensen2011topology}. The parameters are cycled through the optimization algorithm multiple times, each time subject to a steeper logistic function to force them closer to 0 or 1. This gradual binarization prevents the gradients from diverging during the process. The final parameters must be binary so that the design is manufacturable using standard lithographic techniques. Another key fabrication constraint is the minimum feature size of the structure. To enforce a feature size minimum, we supplied a constraint to the optimization algorithm and used a standard morphological filtering technique \cite{sigmund2007morphology}. The mathematical details of the optimization scheme can be found in Appendix \ref{sec:a1}.

In Figure \ref{fig:1}(d) we demonstrate the focusing ability of a diffractive cascade. In this example, the incident illumination is centered at 10 keV, with an energy bandwidth of 10$^{-4}$, a typical value for synchrotron radiation after passing through a monochromator. The zone plate and elements of the diffractive cascade are each made of 400-nm-thick Au atop a 1-$\mu$m-thick Si$_3$N$_4$ membrane. The diameter of each element is 82 $\mu$m and the focal lengths of both devices are 3.3 cm. The cascade is made of 10 elements, spaced 1 cm apart from one another (see Appendix \ref{sec:a1} for the physical and optimization parameters used throughout). Applying topology optimization to this structure results in the cascade shown in Figure \ref{fig:1}(c). The same region surrounding the center of the element is visualized for both the zone plate and the cascade elements. Only a portion of each structure is shown to allow for visibility of the distinct zones, which are small compared to the size of the entire element. We observe a higher proportion of solid material in the earlier elements of the cascade, with some of the early elements resembling zone plates (which are 50\% solid). The later elements are mostly void, indicating the bias towards a low absorption structure when there is a substantial number of elements. This suggests that the cascade is performing a kind of pre-focusing that approximates a zone plate with the earlier elements and refining the beam shape closer to the focal spot. This cascade can focus the incoming illumination much better than the zone plate when both are restricted to same maximum aspect ratio. In the above example the aspect ratio of all features is bounded at 8, which results in low zone plate performance due to the limited interaction with the X-rays this allows. On the other hand, the cascade's elements create repeated interaction with the illumination. This allows for an increase in peak brightness at the focal spot by 3$\times$ and an increase in focusing efficiency from 3\% to 40\%, as shown in Figure \ref{fig:1}(d). For this X-ray energy, a zone plate would have to be roughly 2 $\mu$m thick to achieve the same performance, implying a maximum aspect ratio of 40.

\begin{figure*}
    \centering
    \includegraphics[scale=0.53]{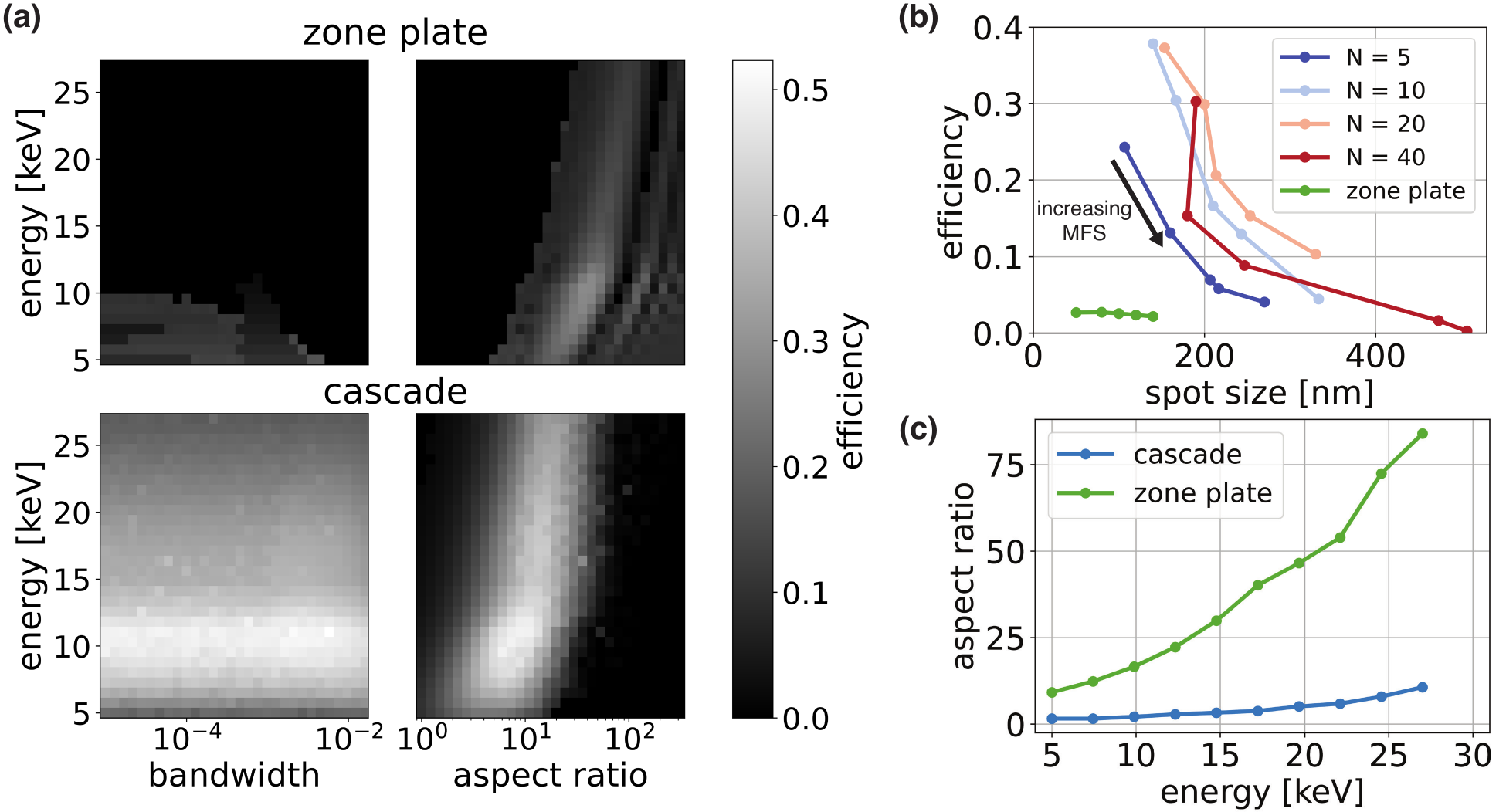}
    \caption{\textbf{Expanded addressable source parameters.} \textbf{(a)} Performance comparison between zone plate and diffractive cascade over a range of parameters. Each point corresponds to the efficiency after optimizing a gold diffractive cascade with $N=10$ elements for the specified source parameters. When sweeping source bandwidth, the aspect ratio is 8. When sweeping aspect ratio, the source bandwidth is $10^{-4}$. \textbf{(b)} Relationship between focal spot size and focusing efficiency. The points on each curve correspond to different minimum feature sizes. Optimal performance is in the top left of the plot where efficiency is high but the spot size is small. The arrow indicates the direction of increasing minimum feature size (MFS). \textbf{(c)} Scaling of the aspect ratio necessary to achieve 10\% efficiency as a function of energy. For each energy the thickness of the elements is increased at a constant minimum feature size until 10\% focusing efficiency is reached.} 
    \label{fig:2}
\end{figure*}

Figure \ref{fig:1}(e) presents a key tradeoff for the diffractive cascade. A cascade with multiple elements is able to outperform a single element due to the increased interaction with the X-rays and the large expressivity enabled by the thousands of parameters being jointly optimized. However, introducing more elements to the cascade also increases the absorption of the X-rays by the elements and attached membranes. Therefore, for each material we observed an optimal number of elements in the cascade to maximize the efficiency. For heavier, more X-ray-absorptive materials this optimum comes earlier, while lighter elements peak at larger values of $N$. There is a wide range of $N$-values over which the efficiency of the cascades exceed that of the zone plate. Because the illumination is normally incident and the index contrast in the X-ray regime is on the order of $10^{-5}$, reflection from the elements is negligible and not considered here.

\section{Expanded addressable source parameters}
One of the biggest limitations of analytic or semi-analytic zone plate designs is the small range of sources they can be used with. That is, the characteristics of the source have to conform to strict requirements to achieve appreciable focusing. This in turn limits the type of measurements and samples that can be accommodated by systems with zone plates. Two of the most relevant source parameters affecting zone plate performance are source energy and bandwidth. Higher energy sources enable increased sample penetration, allowing thicker and more dense samples to be imaged. Higher bandwidth sources deliver more X-ray flux to the sample and shorten exposure times, as well as allow for a range of material interactions that can reveal energy-resolved information. Due to the huge design space that inverse design opens up, a wide range of sources can be accommodated by diffractive cascades. As shown in Figure \ref{fig:2}(a), diffractive cascades achieve substantial focusing at high X-ray energies without requiring large aspect ratios. In Figure \ref{fig:2}(a) each point on the cascade plots represents a separate cascade optimized for that combination of parameters. Above 15 keV, aspect ratios exceeding 50 are required to achieve focusing with a zone plate, while a cascade can focus such energies with an aspect ratio less than 10. In fact, lower aspect ratios make the optimization of diffractive cascades easier. High aspect ratio elements cause sharp changes in gradients during optimization, making discovery of an optimal design more challenging. 

Figure \ref{fig:2}(a) also demonstrates the increased bandwidth tolerable by diffractive cascades. Zone plates are designed for a specific wavelength as per Eq. (\ref{eq:1}), so their performance degrades when the source has a nonzero bandwidth. In practice, a monochromator must be placed before the zone plate to reduce the energy bandwidth below 10$^{-4}$ \cite{caciuffo1987monochromators}. By selecting a specific wavelength, many of the beam photons are discarded. This means fewer photons are available to strike the sample, lengthening exposure times. Therefore, tolerating a wider bandwidth allows for shorter exposure times when imaging. This enables the visualization of short timescale processes that are impossible to image with a highly monochromatic beam. For example, a broadband X-ray beam has been used to image enzyme processes occurring over picosecond timescales \cite{meents2017pink}. A diffractive focusing optic amenable to a high bandwidth beam would enable ultrafast X-ray imaging while maintaining nanoscale resolution. 

\begin{figure*}
    \centering
    \includegraphics[scale=0.49]{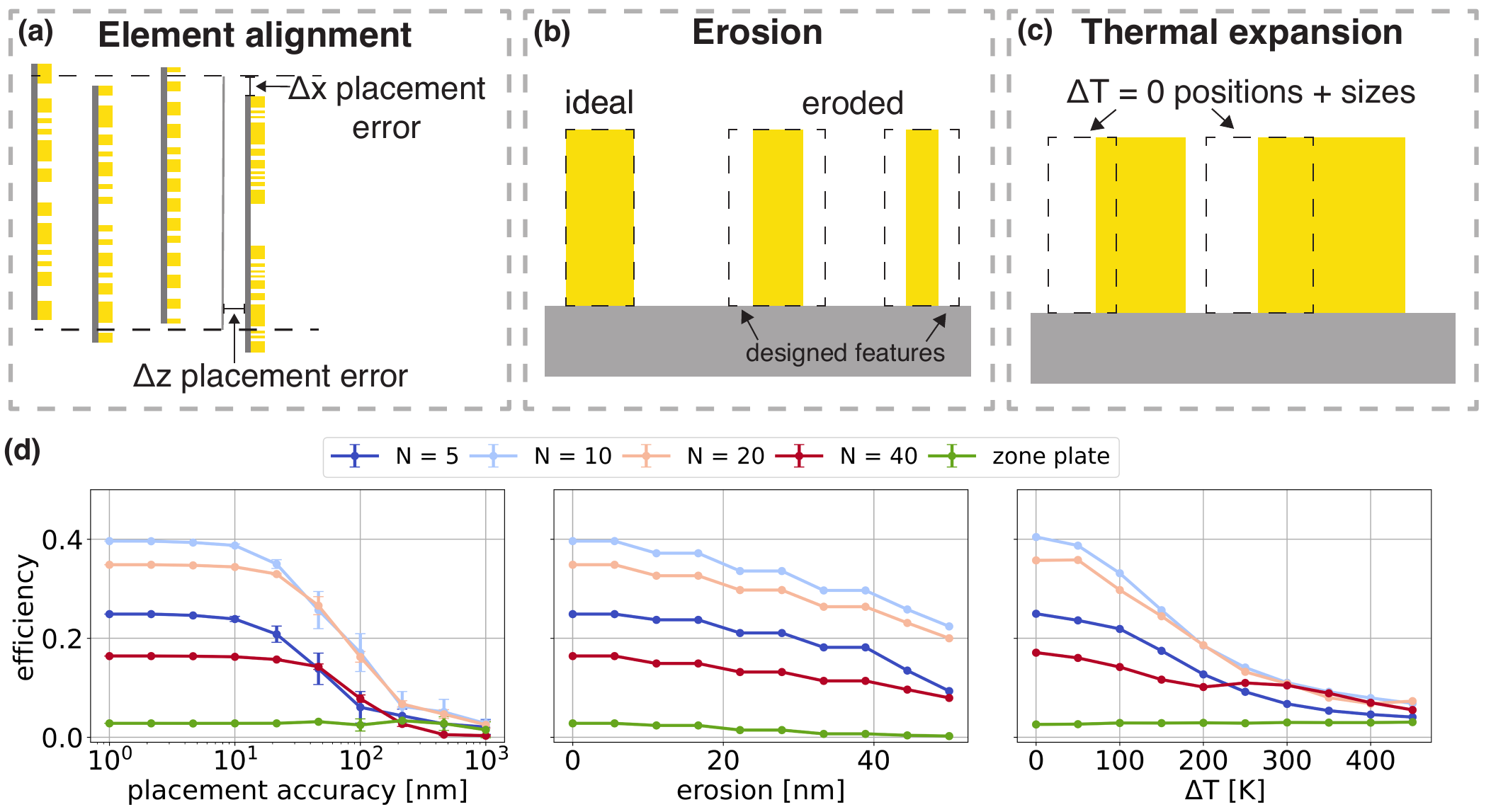}
    \caption{\textbf{Diffractive cascade robustness to perturbations.} \textbf{(a)} Schematic depiction of diffractive cascade element alignment errors. Elements can be misaligned in both the transverse and longitudinal directions. \textbf{(b)} Schematic of the effect of erosion on zone shapes. Erosion refers to over-etching in the lateral direction during etching. \textbf{(c)} Schematic diagram of the effect of thermal expansion on zone shapes. Thermal expansion simultaneously shifts the zones and makes them larger. \textbf{(d)} Effect of each error on focusing performance. Placement accuracy refers to the average displacement from nominal placement both along the optical axis and transverse to it. For each value of the placement accuracy, multiple trials are conducted with random displacements of each element and averaged to obtain the plotted curves. Erosion refers to the amount by which each feature is eroded or over-etched. The thermal response for each material is simulated using the coefficient of thermal expansion at room temperature for that material. The expansion of the membrane and zones are considered simultaneously. $\Delta$T refers to the change in temperature above 300 K.} 
    \label{fig:3}
\end{figure*}

Due to their short wavelengths, X-rays have the opportunity to image smaller features than longer wavelength illumination. Nevertheless, nanoscale resolutions are only possible if the beam can be focused tightly. In Figure \ref{fig:2}(b) we show how diffractive cascades can achieve high efficiency even at small focal spot sizes. We plot the focusing efficiency as a function of the focal spot size at a fixed maximum aspect ratio of 8. Here, the focal spot size is defined as the distance from the center of the focal spot to the first null in the intensity pattern. The spot size is principally controlled by the minimum feature size of the element, so this parameter is swept to generate the curves in Figure \ref{fig:2}(b). Optimal performance corresponds to the top left of the plot, where efficiency is high and the spot size is small. We observe that diffractive cascades can achieve nanoscale focusing at a much higher efficiency than a zone plate. While diffractive cascades do not reach the diffraction-limited focal spot size that zone plates do, they vastly improve efficiency while retaining a nanoscale focal spot. Figure \ref{fig:2}(b) also indicates that while cascades with fewer than 10 elements can focus more tightly, they do so at a lower overall efficiency. This suggests a tradeoff between imaging resolution and diffraction efficiency that can be tuned by changing the number of elements in the cascade. In addition to this tradeoff between efficiency and spot size, topology optimization also allows us to change the size of the area in which to compute the objective function. This enables the tailoring of the spot size based on the needs of the downstream measurement. For example, if nanoscale resolution is not necessary, one can widen the objective area and design a cascade to maximize efficiency. Standard zone plates do not have this flexibility.  

Diffractive cascades also have better aspect ratio scaling characteristics as a function of the source energy. One of the largest limitations of zone plates is the strong scaling with energy of the required aspect ratio. In Figure \ref{fig:2}(c) we plot the aspect ratio required to achieve 10\% focusing efficiency as a function of the incident energy for both a zone plate and a diffractive cascade with 10 elements. We see that the necessary zone plate aspect ratio increases sharply. On the other hand, the diffractive cascades are able to achieve focusing at much lower aspect ratios. Due to the increased optical thickness of the cascades, focusing is not limited by interaction with the X-rays. This means that the aspect ratio necessary to focus hard X-rays is low and only slightly higher than the aspect ratio for soft X-rays. This greatly simplifies the fabrication of hard X-ray optics by allowing the use of standard etching techniques. 

\section{Robustness to placement and fabrication errors, thermal expansion}
During fabrication, alignment, and operation of an imaging system, there are numerous nonidealities that affect the performance of the focusing optic. In this section, we study the impact of three such nonidealities that occur in the use of X-ray optics in beamlines. We demonstrate that diffractive cascades are expected to be robust to these errors and exceed the performance of zone plates even for considerable deviations from an ideal design.

The effect of the three types of perturbations studied are shown in Figure \ref{fig:3}. In Figure \ref{fig:3}(a)--(c) we provide schematic depictions of these three perturbations. Figure \ref{fig:3}(d) shows the quantitative behavior of focusing performance as a function of the severity of these errors. We first consider the effect of element misalignments on the performance of a zone plate and diffractive cascades with different numbers of elements. Specifically, we consider errors along the optical axis as well as within the plane of each element. For each value of the placement accuracy, we randomly displace each element according to a uniform distribution with that mean. For each level of placement inaccuracy, we run multiple trials to obtain the mean values and associated error bars shown in the plot. We observe that the cascades are largely robust to alignment errors up to 10 nm, with performance exceeding that of the zone plate baseline up to 100 nm errors. Alignment within 10 nm is achievable using modern piezoelectric stages, suggesting the diffractive cascades would be robust to the expected misalignment errors in a typical beamline \cite{bajt2018x}. 

Another source of error is the fabrication process of the elements. Metal X-ray zone plates are typically fabricated by etching trenches to create void zones, followed by electroplating of the zone material on the solid regions. However, the etching procedure is imperfect and often results in over- or under-etching. In inverse design these are known as dilation and erosion, respectively. These imperfections result in degraded performance due to the deviation from the expected shape of the zones. To simulate the effect of this, after an optimization of the cascade we performed an erosion operation on the parameters to mimic the effect of over-etching as depicted in Figure \ref{fig:3}(b). By varying the length scale of this erosion we can determine the tolerance of the cascade to this type of fabrication error. We observe that the cascades can comfortably tolerate erosion on the order of 5-10 nm, which is equal to the precision possible with modern etching techniques \cite{gorelick2010direct}. The performance of the cascades stays above that of the zone plate even for extreme levels of erosion. We also observed that cascades have similar robustness to dilation, or under-etching, as shown in Appendix \ref{sec:dilation-sidewall} along with an investigation of sidewall smearing. Taken together, these facts mean that diffractive cascade focusing performance is expected to be robust to common fabrication defects. 

We also studied the effect of heating on the performance of the cascades. During pulsed or continuous operation with extremely bright X-ray sources, the temperature of X-ray optics can rise by as much as 300 K from their idle temperature \cite{nilsson2010simulation}. This heating can cause mechanical instabilities as well as reduce the optical performance of the device due to expansion of the constituent materials. We simulate the effect of this thermal expansion here. A typical zone plate is made of metal zones placed on top of a membrane made of Si$_3$N$_4$. In this simulation, we considered the thermal expansion of both the membrane and the zones themselves. The transverse expansion of the membrane causes movement in the positions of the zones, while the transverse expansion of the zone material causes changes in the shape of the zones. Expansion along the optical axis results in slightly larger thicknesses for the membrane and the zones. Each of these corresponds to a deviation from the optimized design, reducing performance. The effect of thermal expansion is shown in Figure \ref{fig:3}(d). Performance is impacted for all values of $N$, with the most efficient designs maintaining efficiency above 20\% even for the most harsh heating conditions. This indicates diffractive cascades can withstand the thermal conditions within high power X-ray beamlines, enabling their use in short timescale pulsed experiments as well as long exposure measurements. Note that our analysis ignores the potential impact of thermal stress at the Au-Si$_3$N$_4$ interfaces in both the zone plate and cascade elements, which in practice may lead to delamination under extreme conditions.

\section{Multi-plane optimization for extended depth of focus}
\begin{figure}
    \includegraphics[scale=0.67]{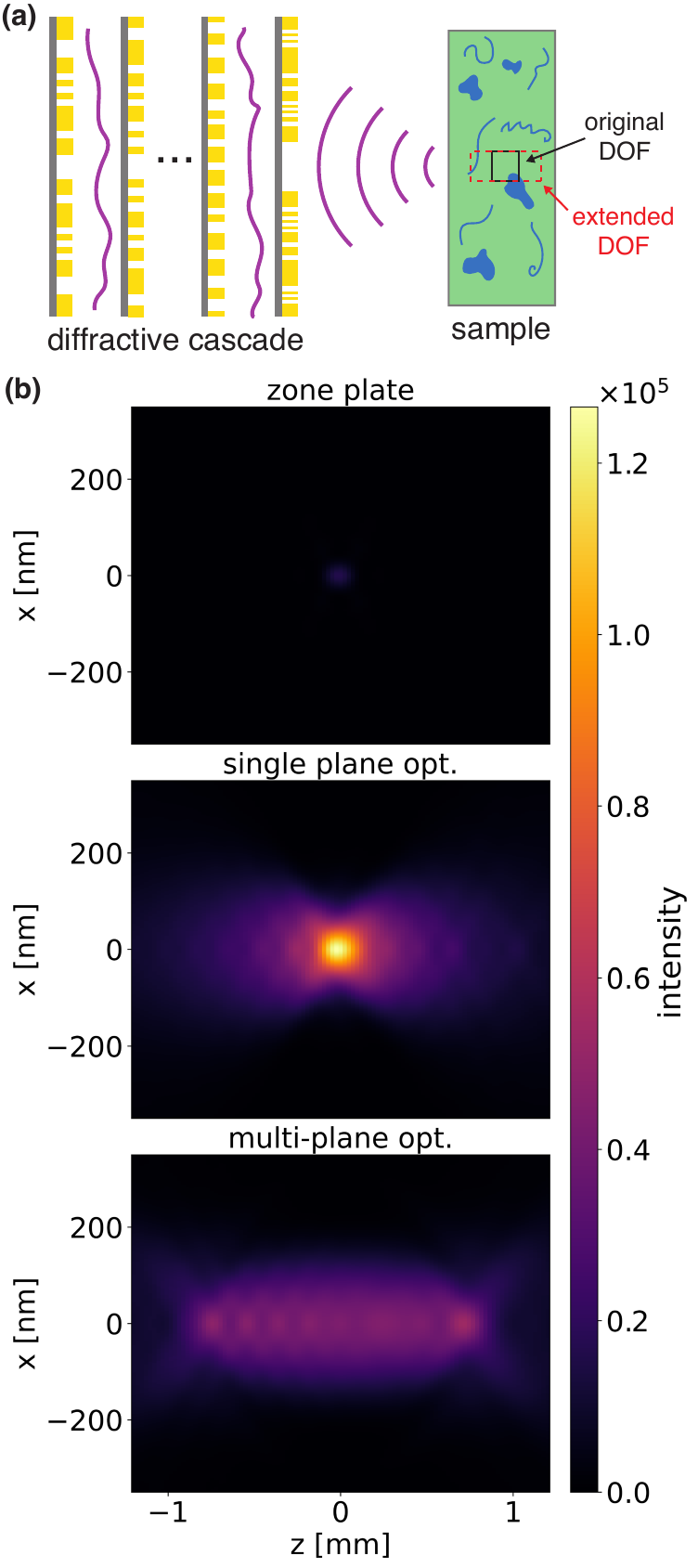}
    \caption{\textbf{Extended depth of focus via multi-plane optimization.} \textbf{(a)} Schematic diagram of extended depth of focus (DOF) imaging using a diffractive cascade. When imaging an extended volume sample, an extended depth of focus can enable simultaneous imaging of a higher proportion of the volume. \textbf{(b)} Comparison of depth of focus of a zone plate, single-plane optimized cascade, and a multi-plane optimized cascade. Optimizing across multiple planes allows for an extended depth of focus at somewhat reduced peak intensity. The x axis is a radial slice transverse to the optical axis. The z axis is the optical axis, with z = 0 referring to the focal plane chosen for the optimization. The results here are for a gold diffractive cascade with 40 elements of diameter 82 $\mu$m with a focal length of 3.3 cm. The aspect ratio is constrained to a maximum of 8.} 
    \label{fig:4}
\end{figure}

Across wavelengths, extending the depth of focus of lenses is useful for applications such as microscopy, tomography, and lithography \cite{kohnen2019visual, lorenser2012ultrathin, yin2017extended, von1992depth}. X-ray measurements can benefit from an extended depth of focus in the same way. In particular, the speed and quality of images taken of extended volumes is often limited by the depth of focus. With a small depth of focus, an imaging system must image the sample in thin slices, lengthening the measurement time and creating the possibility of changes in the sample or imaging system during the scan, corrupting the resulting data. Therefore, oftentimes it is advantageous to have as much of the sample in focus at a time as possible. 

Zone plates have a limited depth of focus that is inherent to the structure of the zone plate itself. The depth of focus of zone plates has only been extended in the visible wavelength regime where zone plate performance is much less constrained \cite{torcal2022sector, trung2024design}. Furthermore, these techniques for extending the depth of focus are tailored specifically to this task, meaning they do not generalize to improving other performance metrics. Topology optimization is well equipped for extending the depth of focus due to the relative ease in altering the objective function. Here, we extend the depth of focus of our diffractive cascades by introducing a multi-plane objective function. Instead of optimizing the power only at the focal plane of the optic, we include terms in the objective function for multiple planes in front of and behind the focal plane. This allows the optimization algorithm to find a design that focuses the incoming beam tightly while also creating a large depth of focus. 

In Figure \ref{fig:4} we demonstrate the concept of an extended depth of focus and the ability for our optimization technique to improve this metric. In Figure \ref{fig:4}(a) we depict how an extended depth of focus enables the imaging of a larger volume in a single exposure. In Figure \ref{fig:4}(b) we compare the depth of focus achieved by a zone plate with those resulting from cascades optimized using single-plane and multi-plane objective functions. With a single-plane in the objective function, the focal spot changes in shape and intensity along the depth of the sample. With multiple planes in the objective function, the focal spot becomes uniform across many depths, indicating an extended depth of focus. For the results shown in Figure \ref{fig:4}, the depth of focus is extended to over three times the value achieved using single-plane optimization. Despite this extension in the depth of focus the spot size in the transverse direction is only expanded by $\sim$50\%. This indicates an ability to achieve nanometer-scale imaging over a range of nearly 2 mm along the optical axis. When comparing to the zone plate, we observe that the depth of focus is greatly extended with much more light in the focal spot.  

Empirically, we noticed a tradeoff between depth of focus and the peak intensity of the focal spot. That is, the highest peak intensity is achievable using single-plane optimization, but the depth of the focus is minimal. Introducing contributions from other planes to the objective function extends the depth of focus, but lowers the peak intensity. Furthermore, we noted a preference for a larger number of elements in the cascade for the multi-plane objective function. For a single-plane objective function, a gold cascade achieves maximum efficiency with $\sim$10 elements. When including multiple planes, 40 elements is optimal. This is due to the increase in complexity caused by including multiple planes in the objective function. This increased complexity shifts the tradeoff between absorption and expressivity towards favoring more parameters. The focal spot shape along the optical axis was most uniform when placing the objective function planes $\sim$70 $\mu$m from one another. Larger spacing led to discrete focal spots with empty gaps between them.

\section{Discussion}
In this work we have demonstrated the application of photonic topology optimization to the design of diffractive cascades for X-ray focusing. In simulation, these cascades show the ability to focus X-rays across a wider range of source parameters than a traditional zone plate. This focusing is more efficient and achievable at a smaller aspect ratio, allowing high quality X-ray focusing to be accomplished with less difficult fabrication techniques. These cascades are also robust to practical perturbations such as misalignment, fabrication errors, and thermal expansion due to excessive illumination. We also showed how the objective function can be modified to facilitate the enhancement of metrics such as the depth of focus. 

Improvements in X-ray optics have historically been driven by advances in materials science and fabrication. Due to the low volume and bespoke nature of X-ray optics, this has limited increases in key figures of merit. At the same time, the X-ray wavelength regime has been overlooked by the inverse design community due to the huge computational resources required to do full 3D simulations of X-ray propagation. This work demonstrates the ability for a few key assumptions to unlock substantial improvement in X-ray optics via optimization. Beyond focusing, this work could be extended to create optical elements performing a wide range of X-ray manipulation tasks. In particular, many applications would be enabled with the relaxation of the cylindrical symmetry constraint. Because nanofocusing X-ray optics require minimum feature sizes on the order of tens of nanometers, performing full 3D scalar diffraction simulations is infeasible for structures more than hundreds of microns in diameter. In particular, such simulations are bounded by GPU memory and the speed of FFTs during wave propagation. If full 3D simulations were made tractable, objective functions and elements without cylindrical symmetry could be considered. This would unlock applications such as projection lithography, beam steering, and edge detection. 

Another open question is the applicability of the methods described here to X-ray sources without full spatial coherence. Modern free-electron lasers have spatial coherence lengths approaching the full width of the beam, meaning such radiation is well described by fully spatially coherent plane waves \cite{geloni2010coherence}. However, less advanced X-ray sources do not have these coherence properties, with extremely incoherent sources such as X-ray tubes having coherence lengths on the order of nanometers. In Appendix \ref{sec:a2} we study the effect of a partially coherent source on the focusing efficiency of diffractive cascades. This analysis suggests our design method suffers little performance degradation for moderately incoherent sources. The ability to design X-ray optics for partially coherent X-rays would impact multiple applications. Specifically, it could enable the imaging and lithography applications discussed here without the need for an expensive and complicated beamline facility. This would have the potential to enable the translation of sophisticated X-ray imaging and characterization techniques from large synchrotron facilities to the benchtop. Most lab-scale X-ray experiments utilize relatively small, partially coherent X-ray sources such as X-ray tubes. The ability to perform experiments that are currently relegated to complex shared-user synchrotrons in more conventional lab settings could accelerate discovery and make advanced characterization routine.

As discussed earlier, the aspect ratio requirements for efficient hard X-ray zone plates are restrictive. This has resulted in the development of advanced fabrication methods such as metal-assisted chemical etching that have enabled zone plate aspect ratios of up to 500 \cite{chang2014ultra, li2017fabrication}. While these methods increase focusing performance, they are highly specialized and difficult to implement without substantial fabrication expertise and process tuning. In this work, we demonstrated the focusing of X-rays with low aspect ratio structures that are unlikely to require highly specialized processing. Even for X-ray energies exceeding 20 keV, the required aspect ratio for focusing is less than 10. Gold zone plates with such aspect ratios have been fabricated for decades with simple techniques such as electron beam lithography \cite{shaver1978x}. These methods have also been refined to produce zone plates with aspect ratios exceeding 10 \cite{divan2002progress}. Due to the widespread diffusion and utilization of these standardized cleanroom techniques, our results point toward the inexpensive and widespread deployment of efficient diffractive X-ray lenses. The elements of our diffractive cascades retain the basic circular structure of zone plates, meaning many of the previously developed fabrication methods are readily transferable.  

\section{Acknowledgments}
W.M. and J.C. acknowledge support from the National Science Foundation Graduate Research Fellowship Program under grant number 2141064. S.P. gratefully acknowledges support from the MathWorks Engineering Fellowship via MIT. C.R.C. is supported by a Stanford Science Fellowship. The authors acknowledge MIT SuperCloud and the Lincoln Laboratory Supercomputing Center for providing high-performance computing resources. This work was funded in part by the U. S. Army Research Office through the Institute for Soldier Nanotechnologies (ISN) at MIT, under Collaborative Agreement Number W911NF2320121. We thank Yaocheng Tian and Colin Gilgenbach for helpful discussions. The authors used generative AI tools for codebase development. 
\bibliography{main}

\clearpage
\onecolumngrid
\appendix

\begin{center}
    \textbf{\large Appendix}
\end{center}
\section{Optimization details}
\setcounter{figure}{0} 
\renewcommand{\thefigure}{A.\arabic{figure}} 
\label{sec:a1}
In this section we specify the mathematical details of the optimization procedure. We also provide the physical and computational parameters used in the experiments presented in the main text. We first define the optimization problem in question as
\begin{align}
    & \min_{\boldsymbol{\rho}} \mathcal{L}(\boldsymbol{\rho}) \\
    \text{subject to: }   & g_{\text{solid}}(\Bar{\boldsymbol{\rho}}) \leq 0 \\
    & g_{\text{void}}(\Bar{\boldsymbol{\rho}}) \leq 0 \\
    & 0 \leq \rho_l(r) \leq 1 \quad \forall l, r
\end{align}
Where $\boldsymbol{\rho} = \{\rho_l(r)\}_{l=1}^N$ is the set of radial density profiles for each of the $N$ elements. The forward model consists of interleaved propagation through free space and the application of the cascade elements. The application of an element transforms the incoming wave according to the thin element approximation
\begin{equation}
    U_l^+(r, \lambda) = U_{l-1}^-(r, \lambda) \cdot t(\Bar{\rho}_l(r), \lambda),
\end{equation}
where $t(\Bar{\rho}_l(r), \lambda)$ is the complex transmission function of the $l$-th element at the wavelength $\lambda$. The wavefront incident on the next element is determined using the angular spectrum propagation kernel 
\begin{equation}
    H(f_x, f_y) = \exp\Big[i2\pi \frac{z}{\lambda}\sqrt{1 - (\lambda f_x)^2 - (\lambda f_y)^2}\Big].
\end{equation}
With this the expression for the field after propagation is
\begin{equation}
    U_l^-(r, \lambda) = \mathcal{F}^{-1}\{H(f_x, f_y)\cdot \mathcal{F}\{U_l^+(r, \lambda)\}\},
\end{equation}
where $\mathcal{F}\{\cdot\}$ is the Fourier transform. The objective function is computed as the integrated power within a small region $\Omega$ around the center of the focal plane
\begin{equation}
    \mathcal{L}(\boldsymbol{\rho}) = - \sum_\lambda \int_\Omega |U_N^-(r, \lambda)|^2.
\end{equation}
The forward model is evaluated on the parameters after they have been filtered and projected. The filter function filters out features below a given size. It is defined as, 
\begin{equation}
    f_{\text{filter}}(\rho_l(r)) = \Tilde{\rho}_l(r) = \int K(r-r') \rho_l(r') dr',
\end{equation}
where $K(r)$ is the spatial kernel
\begin{equation}
    K(r) = \frac{w(r)}{\int w(r') dr'}.
\end{equation}
$w(r) = \max(0, R - |r|)$ is the weighting function with $R$ being the desired minimum feature size. $\Bar{\rho}_l(r)$ is obtained by applying a threshold function to the filtered parameters
\begin{equation}
    \Bar{\rho}_l(r) = f_{\text{thresh}}(\Tilde{\rho}_l(r); \beta) = \frac{\tanh(\beta / 2) + \tanh[\beta(\Tilde{\rho}_l(r)-1/2)]}{2\tanh(\beta / 2)},
\end{equation}
where $\beta$ controls the strength of the projection. To further ensure that no features below the minimum are included in the final design, we supply inequality constraints to the optimization algorithm. These constraint functions are computed for each element and then aggregated together. For a single element the feature size constraints are computed as
\begin{align}
    g_{s,l} &= |\max(0, \Bar{\rho}_l(r) - \mathcal{O}(\Bar{\rho}_l(r))| - \epsilon \\ 
    g_{v,l} &= |\max(0, \mathcal{C}(\Bar{\rho}_l(r))- \Bar{\rho}_l(r)| - \epsilon
\end{align}
Where $\epsilon$ is the constraint violation tolerance.  $\mathcal{O}(\cdot)$ and $\mathcal{C}(\cdot)$ are the smooth opening and smooth closing operations as defined in \cite{sigmund2007morphology}. These per-element constraint functions are then aggregated into a single number for the solid and void regions  respectively
\begin{align}
    g_{\text{solid}}(\Bar{\boldsymbol{\rho}}) &= \frac{1}{\beta_{\text{agg}}} \ln \left(\sum_{l=1}^N \exp(\beta_{\text{agg}}\cdot g_{s,l})\right), \\ 
    g_{\text{void}}(\Bar{\boldsymbol{\rho}}) &= \frac{1}{\beta_{\text{agg}}} \ln \left(\sum_{l=1}^N \exp(\beta_{\text{agg}}\cdot g_{v,l})\right).
\end{align}
Below we list the parameters used for the forward model and the optimization procedure. These parameters are used for all the results in the main text except where otherwise noted. For example, in the plots where we show cascade performance as a function of source bandwidth and energy, these parameters vary while all other parameters take the values in the tables below. The energy-dependent refractive indices of the materials considered in this work are taken from \cite{henke1993x}. 
\begin{table}[!htbp]
    \centering
    \begin{minipage}[t]{0.48\textwidth}
        \centering
        \begin{tabular}{lc}
        \toprule
        \textbf{Parameter} & \textbf{Value} \\
        \midrule
        Central X-ray energy & 10 keV \\
        Source energy bandwidth & 10$^{-4}$ \\ 
        Minimum feature size & 50 nm \\
        Number of elements & 10 \\
        Source wavelengths & 11 \\
        Grid spacing & 10 nm \\
        Element diameter & 82 $\mu$m\\
        Zone thickness & 400 nm\\
        Membrane thickness & 1 $\mu$m \\
        Inter-element distance & 1 cm \\
        Focal length & 33 mm \\
        \bottomrule
        \end{tabular}
        \caption{Physical parameters.}
        \label{tab:physical_params}
    \end{minipage}
    \hfill 
    \begin{minipage}[t]{0.48\textwidth}
        \centering
        \begin{tabular}{lc}
        \toprule
        \textbf{Parameter} & \textbf{Value} \\
        \midrule
        Maximum iterations & 100 \\
        $\beta_{\text{min}}$ & 1 \\
        $\beta_{\text{max}}$ & 256 \\
        $\beta_{\text{agg}}$ & 10 \\
        $\epsilon$ & 10$^{-6}$ \\
        $\Omega$ diameter & 24 nm \\
        \bottomrule
        \end{tabular}
        \caption{Optimization parameters.}
        \label{tab:opt_params}
    \end{minipage}
\end{table}
\section{Effect of partial spatial coherence}
\label{sec:a2}
While modern synchrotron and free-electron laser sources are highly coherent, full spatial coherence is still an imperfect assumption. In this section we introduce a method for modeling partial spatial coherence and apply it to diffractive cascades. 

\begin{figure}[h!]
    \centering
    \includegraphics[scale=0.5]{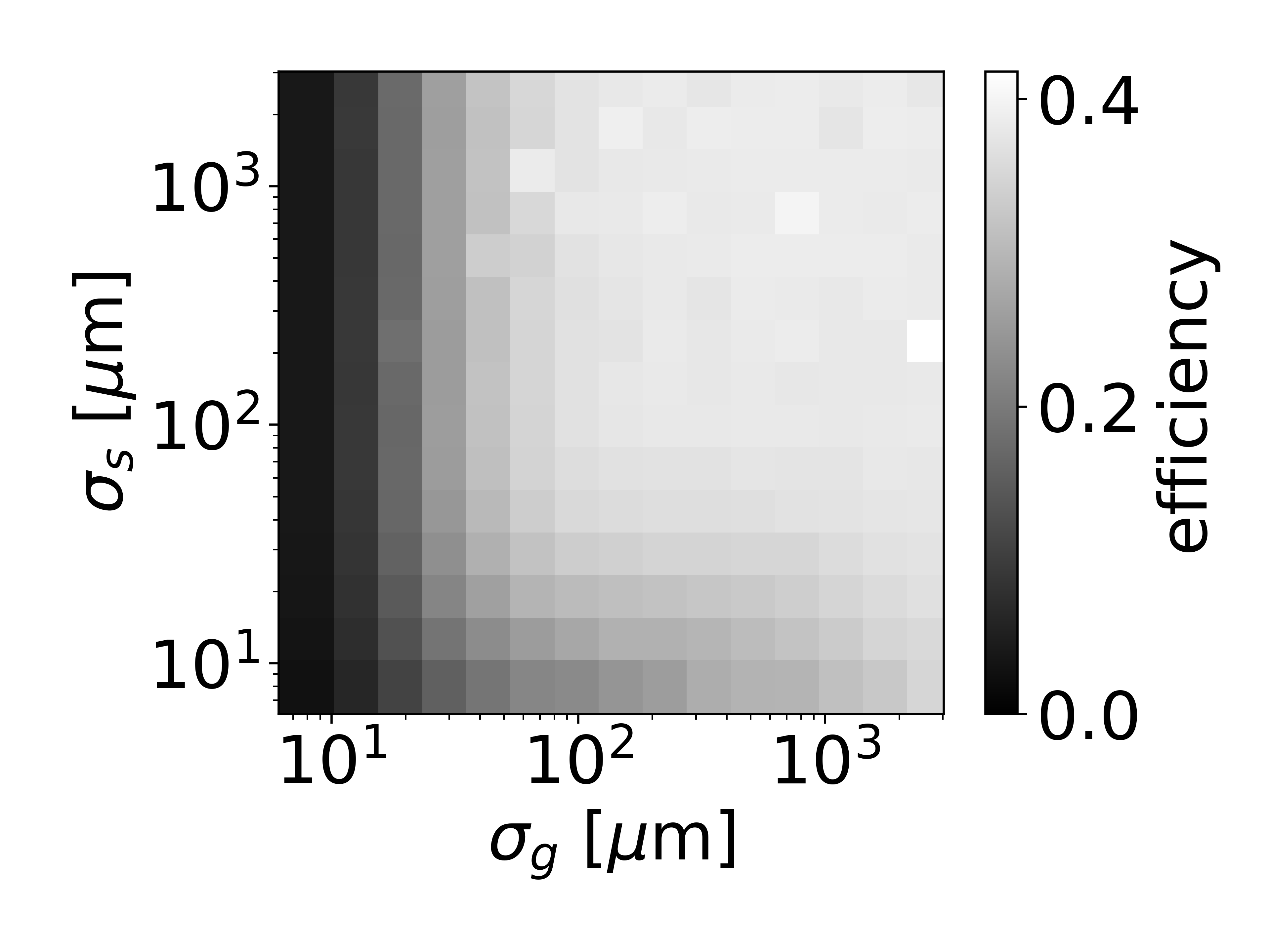}
    \vspace{-0.2cm}
    \caption{\textbf{Effect of partial coherence on cascade performance.} Plot of the efficiency of cascades designed for different degrees of partial spatial coherence. For each combination of $\sigma_s$ and $\sigma_g$ a cascade is designed with the corresponding partially coherent beam in the forward model. The physical and optimization parameters are those specified in Tables \ref{tab:physical_params} and \ref{tab:opt_params}. For the fully coherent model, the cascade with these parameters achieved a focusing efficiency of 41\%.} 
    \label{fig:a1}
\end{figure}

Previous studies of partially coherent X-ray radiation have utilized Monte Carlo simulations \cite{salditt2011partially}. However, a more computationally elegant method uses the cross spectral density. The cross spectral density quantifies the correlation between pairs of points on a wavefront \cite{goodman2015statistical}. A standard form of the cross spectral density is the Gaussian-Schell model which we write here in polar coordinates
\begin{equation}
    W(r_1, \phi_1, r_2, \phi_2) = \exp\left[-\left(\frac{1}{4\sigma_s^2} + \frac{1}{2\sigma_g^2}\right)(r_1^2+r_2^2)\right]\exp\left[\frac{r_1r_2\cos(\phi_1-\phi_2)}{\sigma_g^2}\right],
\end{equation}
where $\sigma_s$ is the beam width and $\sigma_g$ is the coherence length. For efficient computation it is convenient to expand the cross spectral density into coherent modes
\begin{equation}
    W(r_1, \phi_1, r_2, \phi_2) = \sum_{l=-\infty}^{\infty}\sum_{p=0}^{\infty} w_{lp} \Psi_{lp}(r_1, \phi_1)\Psi^*_{lp}(r_2, \phi_2),
\end{equation}
where $\Psi_{lp}(r, \phi)$ are the Laguerre-Gaussian modes and $w_{lp}$ their corresponding weights. Since the modes are mutually incoherent we can propagate each of them individually and combine their results at the end. To determine the effect of partial coherence on the performance of our diffractive cascades, we replaced the fully spatially coherent illumination used in the main text with a Gaussian-Schell model beam. In particular, for each combination of $\sigma_s$ and $\sigma_g$ we computed the coherent mode decomposition of the Gaussian-Schell model. We then truncated the decomposition where $w_{lp}$ dropped below $10^{-3}$. This ensures that the decomposition faithfully represents the true coherence properties of the beam. Figure \ref{fig:a1} shows the focusing efficiency of diffractive cascades designed for different levels of spatial coherence. The fully coherent limit is in the top right of the plot and corresponds to the limits $\sigma_s/D, \sigma_g/D \gg 1$, where $D$ is the diameter of a cascade element. We observe that diffractive cascades maintain their performance even for modest spatial coherence of the beam. 

The efficient performance of diffractive cascades under partially coherent illumination suggests they are effective with multiple common X-ray sources. The most coherent X-ray sources are free-electron lasers, which often have spatial coherence lengths exceeding their beam size, resulting is essentially full spatial coherence \cite{gutt2012single}. Slightly less coherent are synchrotron sources which can achieve coherence lengths of 100-200 $\mu$m \cite{chubar2011development}. With further optimization for partial coherence, laboratory-scale sources could also potentially be used with diffractive cascades. Liquid metal jets have small spot sizes and are capable of coherence lengths of roughly 10 $\mu$m \cite{hemberg2003liquid}, just beyond the point where efficiency drops off steeply in Figure \ref{fig:a1}. 

\section{Physical factors governing the optimal number of cascade elements}
In this section we analyze the physical factors that determine the optimal number of elements in a diffractive cascade. The empirical curves in Figure \ref{fig:1}(e) show that cascade performance peaks at a material-dependent value of $N$, then declines. This behavior reflects a competition between two opposing effects: the increased focusing ability afforded by additional diffractive elements, and the increased absorption incurred by the X-rays passing through those elements. We construct a simple model to illustrate how these two effects interact.

We write the figure of merit as a product of two terms,
\begin{equation}
    FOM(N) = T(N)\eta(N).
\end{equation}
$T(N)$ is the transmission of a cascade with $N$ elements, quantifying the loss due to absorption as the X-rays pass through additional cascade elements. $\eta(N) $captures the saturation of focusing performance with increasing $N$: adding elements expands the space of possible designs and improves performance, but with diminishing returns, since each additional element contributes less incremental benefit when the cascade already contains many elements.

Computing $T(N)$ is simply a matter of considering the absorption due to multiple layers of material with a given refractive index. We do not consider reflection between air and the membrane or the membrane and the element material due to the negligible refractive index contrast. The transmission as a function of the number of cascade elements is
\begin{equation}
    T(N) = T_1^N = \exp\left[-N\frac{4\pi}{\lambda}\left(\beta_mt_m + f\beta_zt_z\right)\right],
\end{equation}
where $n_m = 1 - \delta_m +i\beta_m$ and $t_m$ are the refractive index and thickness of the membrane material, respectively. $n_z = 1 - \delta_z +i\beta_z$ and $t_z$ are the refractive index and thickness of the element zones. $f$ is the average fill factor of the elements in the cascade. Because $T(N)$ depends directly on the imaginary part of the refractive index of the zone material, it differs substantially across materials: heavier elements such as gold are far more absorptive than lighter ones such as silicon, causing $T(N)$ to decay more rapidly with $N$. This explains qualitatively why the optimal number of elements is smaller for denser, more absorptive materials.

The saturation of focusing performance with $N$ is modeled as
\begin{equation}
\label{eq:b3}
    \eta(N) = 1 - e^{-kN},
\end{equation}
where $k$ is a rate parameter controlling how quickly performance saturates as elements are added. This functional form reflects the expectation that the marginal benefit of an additional element decreases as the cascade grows. The product $T(N)\eta(N)$ therefore rises initially, as the gain from additional elements outweighs the cost of absorption, and then falls once absorption becomes dominant.

\begin{figure}[h!]
    \centering
    \includegraphics[scale=0.5]{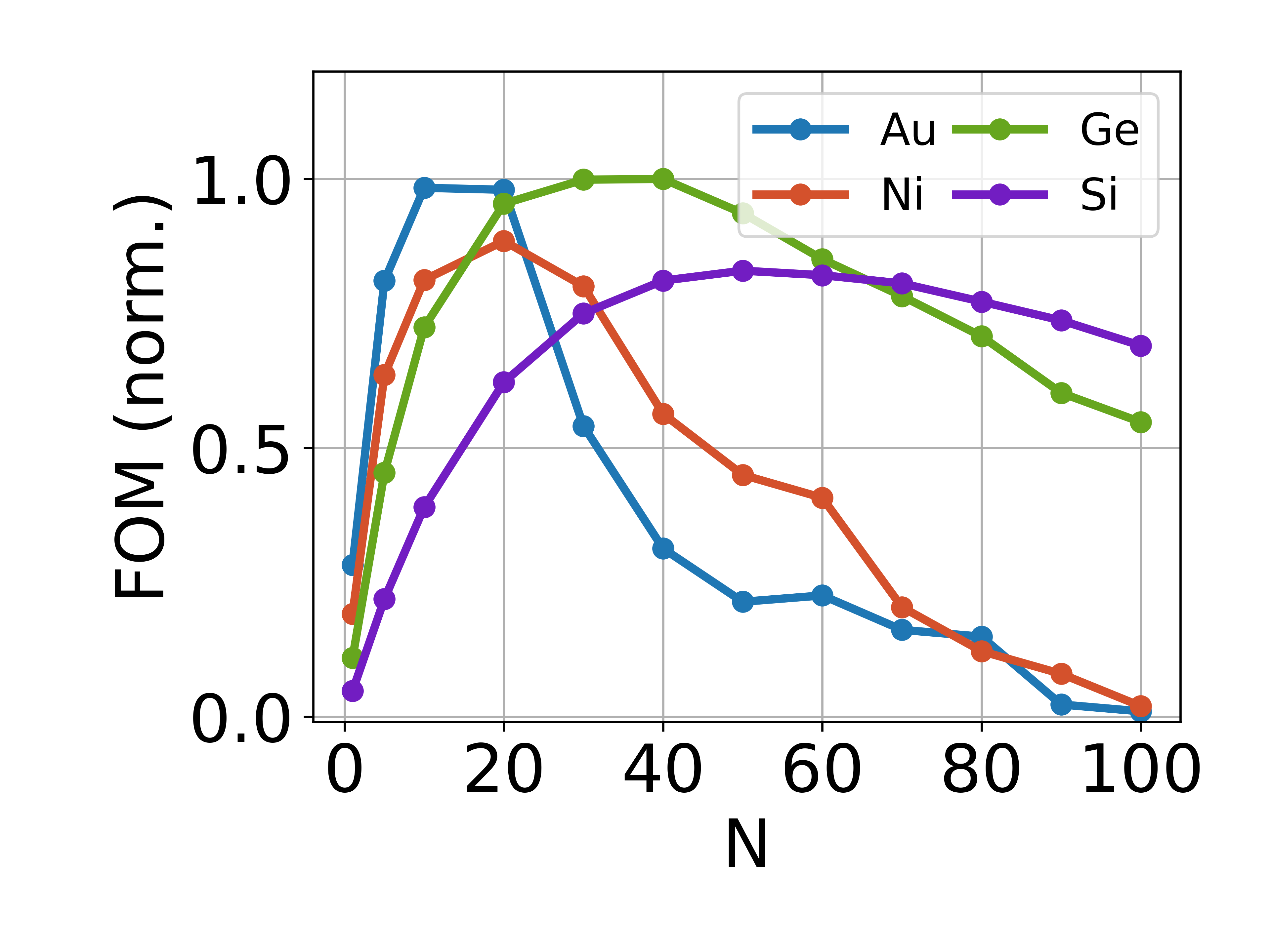}
    \vspace{-0.5cm}
    \caption{\textbf{Cascade element tradeoff figure of merit.} Plot of the figure of merit as a function of the number of elements in the cascade for different zone materials. For each combination of a material and $N$, the fill factor is determined by the corresponding optimization in Figure \ref{fig:1}(e). The learning rate for each material is determined by fitting the $k$ parameter in Eqn. \ref{eq:b3} based on the first three point of the efficiency plot in Figure \ref{fig:1}(e).} 
    \label{fig:a2}
\end{figure}

In Figure \ref{fig:a2} we plot this figure of merit as a function of $N$ for different materials, using the fill factor at each point taken from the corresponding design in Figure \ref{fig:1}(e). The model peaks at roughly the same number of elements as is found to be optimal empirically, and correctly captures the ordering of materials by their optimal $N$. This agreement indicates that the competition between absorption and the saturation of focusing performance is the principal physical mechanism governing the optimal cascade depth, and that material choice enters primarily through its effect on $T(N)$.

\section{Cascade performance as a function of focal length and inter-element distance}
One of the advantages of diffractive cascades is their ability to accommodate a range of focal lengths. In order to change the focal length of a zone plate one must change the zone plate diameter or feature size. Changing the focal length of a diffractive cascade is only a matter of moving the objective region to a different plane.  

Similarly, diffractive cascades can accommodate a range of distances between the elements. While efficiency peaks when the distance between elements is roughly 1 cm, efficiency is high across a range of distances. This allows diffractive cascades to conform to the constraints of the system they are integrated into.  
\begin{figure}[h!]
    \centering
    \includegraphics[scale=0.6]{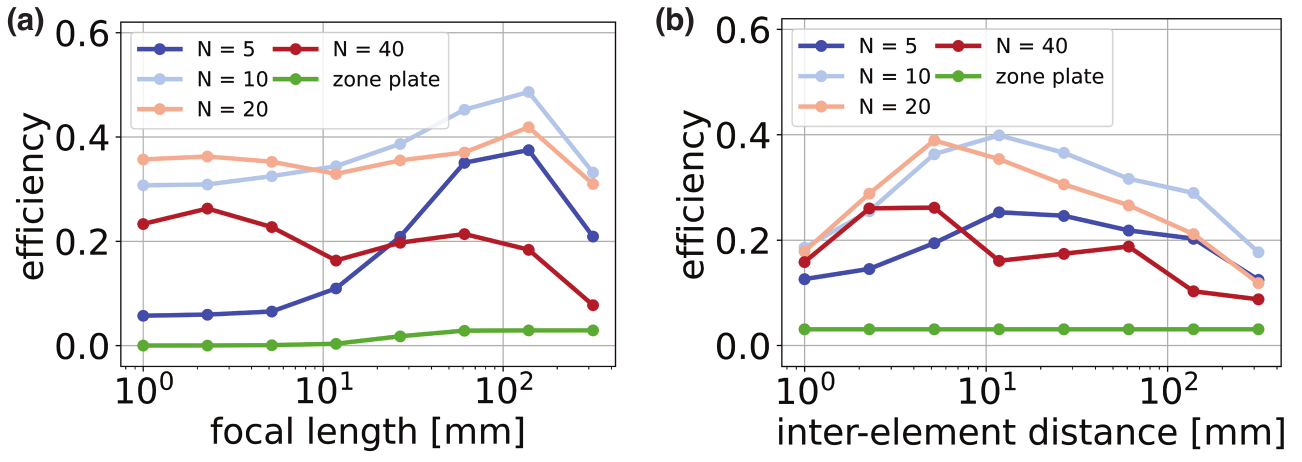}
    \caption{\textbf{Cascade performance as a function of focal length and inter-element distance.} \textbf{(a)} For each focal length a cascade is designed using the specified plane in the objective function. The zone plate focal length is varied by changing the minimum feature size to keep the element diameter constant. \textbf{(b)} For each inter-element distance a distinct cascade is designed. The zone plate design is identical for all values of the inter-element spacing.} 
    \label{fig:a3}
\end{figure}

\section{Robustness of cascades to dilation and transition smoothness}
\label{sec:dilation-sidewall}
\begin{figure}[h!]
    \centering
    \includegraphics[scale=0.53]{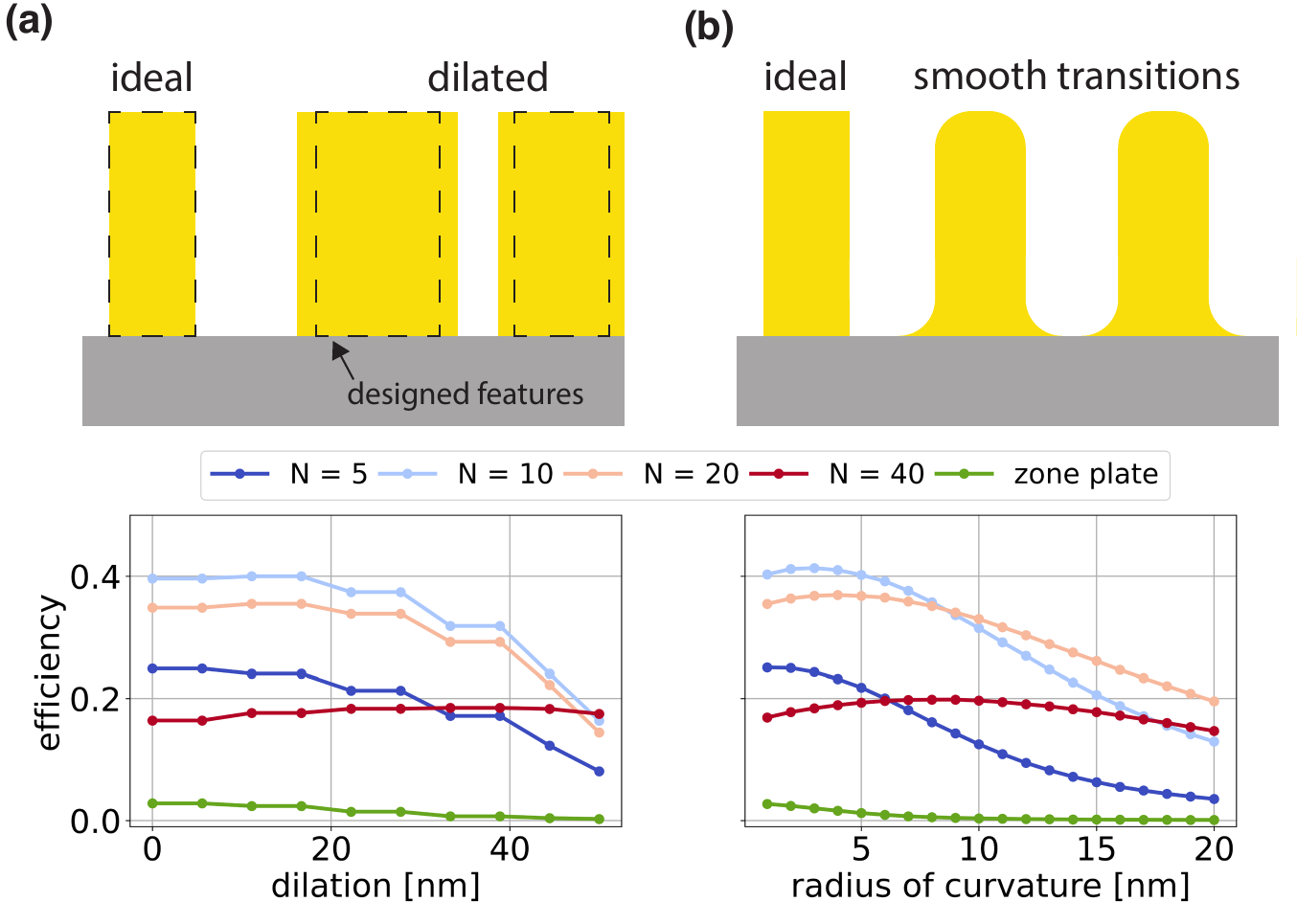}
    \caption{\textbf{Robustness of cascades and zone plates to dilation and sidewall smearing.} \textbf{(a)} Effect of dilation or under-etching on cascade and zone plate performance. Dilation refers to the amount by which each feature is expanded. \textbf{(b)} Effect of smooth solid-void transitions on cascade and zone plate performance. The radius of curvature refers to the radius of the corners of each zone. Zero radius of curvature corresponds to a perfectly abrupt transition.} 
    \label{fig:a4}
\end{figure}

In the main text, we showed the robustness of diffractive cascades to erosion, or over-etching. The complementary phenomena is dilation, or under-etching. Another possible fabrication error is smearing of the transitions between solid and void regions. The effect of both of these imperfections is shown above in Figure \ref{fig:a4}. 

We observe that the cascades are similarly robust to dilation as they are erosion. For cascades with large numbers of elements, some dilation is actually moderately beneficial to performance. Similarly, for some cascades, small amounts of sidewall smearing increase the performance of the system before causing degradation. We conclude that diffractive cascades outperform zone plates even for large amounts of dilation and transition smoothening. 

\end{document}